filename="barrier.TEX"

\input{epsf.sty}
\documentstyle[12pt,]{article}
\begin{document}
\begin{titlepage}
 \hfill{IMAFF-PCC/99-17}
\\
\\
\vskip 2.0cm
\begin{center}
{\large \bf Time of arrival through a quantum barrier\footnote{Work
supported in part by DGESIC PB97-1256.}}
\end{center}
\vskip 1.3  cm
\begin{center}
{\large  J. Le\'on, J. Julve, P. Pitanga}\footnote{ Instituto de
F\'\i sica, UFRJ. 21945-970 Rio de Janeiro- Brasil.} and {\large F.
J. de Urr\'{\i}es}\footnote{ Departamento de F\'{\i}sica,
Universidad de Alcal\'a, Alcal\'a de Henares, Spain.\\ \indent
e-mail addresses: julve@fresno.csic.es, leon@iff.csic.es,
pitanga@if.ufrj.br,\\ \indent fernando.urries@alcala.es }
\\
\vskip .1 cm
  {\sl Instituto de Matem\'aticas y F\'{\i}sica
Fundamental, CSIC}
\\
 Serrano 113-bis, 28006 MADRID, Spain

\end{center}
\date{today}
\vskip 0.5cm

\begin{abstract}
\noindent We introduce a formalism for the calculation of the time
of arrival $t$ at a detector of particles traveling through
interacting environments. We develop a general formulation that
employs quantum canonical transformations from the free to the
interacting cases to compute $t$. We interpret
 our results for the time of arrival operator in terms
of a Positive Operator Valued Measure. We then compute the
probability distribution in the times of arrival at a detector for
those particles that - after their initial preparation - have
undergone quantum tunneling or reflection due to the presence of
potential barriers. We make a extensive analysis of several cases,
the main results being the presence of the expected retardation or
advancement for transmission, and of non-foreseen two bump
structures for some cases of reflection.
\end{abstract}
\vfill{ IMAFF-PCC/99-17}
\newline
\noindent March 1999
\end{titlepage}
\setcounter{footnote}{0} \vskip2truecm
%


\section{Introduction}
In this paper we work out a theoretical framework to compute the
time in which a particle that moves in an interacting environment
arrives at a given point. We apply our results to answer the long
pending question: How long does it take for a particle to tunnel
across (or through) a quantum barrier? In the construction of this
framework we will have to deal with problems of very different kind
that we introduce now:

First, there is the nature of time in quantum mechanics. It appears
as the external evolution parameter in the Schr\"odinger and
Heisenberg equations, common to both, systems and observers alike.
However, time arises in many instances (transitions, decays,
arrivals, etc.) as a property of the physical systems. The attempts
to promote time to the category of observable run early into the
obstruction detected by Pauli~\cite{Pauli}: A self adjoint time
operator implies an unbounded energy spectrum. This was soon
related to the uncertainty relation for time and energy, whose
status and physical meaning has been subject of
controversies~\cite{Mandelstamm,Aharonov1,Fock,Aharonov2}, and is
still subject of elucidation today (see for instance~\cite{Busch1}
and~\cite{Hilgevoord}). The question remains unsettled for closed
quantum systems, specially in the case of quantum gravity, whose
formulation is pervaded by the so called {\sl problem of
time}~\cite{Isham}.

Second, the definition of the time-of-arrival ({\sl toa}), which is
probably the simplest candidate time to become a property of the
(arriving) physical system, rather than a mere external parameter.
Due to its conceptual simplicity, it has been used in many cases to
illustrate different problems related to the role of time in
quantum theory. Allcock analyzed~\cite{Allcock} extensively the
difficulties met by the {\sl toa} concluding they were
insurmountable. The present situation is ambiguous. On the one
side, there are sound theoretical
analysis~\cite{Aharonov3,Oppenheim1} of the {\sl toa} showing that
it can not be precisely defined and measured in quantum mechanics.
This contradicts the possibility of devising high efficiency
absorbers~\cite{Brouard},  that could be used as almost ideal
detectors for {\sl toa}~\cite{Muga5}. On the other hand, there are
explicit constructions of a self-adjoint {\sl toa} operator for the
non-relativistic free particle in one space dimension~\cite{Tate},
the relativistic free particle in 3-D~\cite{Juan}, both avoiding
the Pauli problem. There is also an
 alternative formulation~\cite{Giannitrapani1}  as a
 Positive Operator Valued Measure (POV). Finally, the {\sl toa} has been
 measured in high precision experiments~\cite{Steinberg3, Steinberg4}
 on the arrival of two entangled photons produced by parametric
 down-conversion, one of which has undergone tunneling through a
 photonic band gap (PBG). The experimental results that show superluminal
 tunneling, neatly identify the Hartman effect~\cite{Hartman}
 and the Wigner time delay~\cite{Wigner} (or phase time) as the
 physically relevant mechanisms for the tunneling time and {\sl
 toa} respectively. Whether these results apply only to photons and
 are due to the specific properties of the PBG used, or can be extended
 to other particles and barriers, can not be decided in the lack of a
 satisfactory theory of the {\sl toa} at a space point through interacting
 environments.

The third question is thus the tunneling time, for which there are
three main proposals. Wigner introduced the phase time in his
analysis~\cite{Wigner} of the relationship between retardation,
interaction range, and scattering phase shifts. Buttiker and
Landauer introduced the traversal time~\cite{Buttiker1} in their
study of tunneling through a time-dependent barrier. Soon after,
Buttiker used the Larmor precession as a clock~\cite{Buttiker2},
identifying the dwell~\cite{Smith}, traversal, and reflection times
as three characteristic times describing the interaction of
particles with a barrier. Recent reviews that include these and
other approaches, discussing {\sl toa} and tunneling times from a
modern, unified perspective, can be found in~\cite{Muga4}
and~\cite{Muga6}. The light shed on these questions by the two
photon experiments is revised in~\cite{Chiao1} and~\cite{Chiao3}.

The plan of the work is as follows: In Section 2 we construct the
generic {\sl toa} formalism, generally valid, but still awaiting
the necessary connections to each physical situation of interest.
The starting point is the case of the free particle. A suitable
canonical transformation, similar to that used in scattering
theory, gives the {\sl toa} in interacting environments, (also at
points where $V(x)\neq 0$). In Section 3 we specialize to the case
of potential barriers, showing the existence of two characteristic
times of arrival at each point, one for  incoming and other for
outgoing particles, in the same way as there are two different
kinds of scattering states associated to the free ``in" and ``out"
asymptotic states. In Section 4 we make some calculations and give
estimates of interest for the case of the square barrier. Here we
give the {\sl toa} for both, transmitted and reflected particles,
and discuss some properties of the associated probability
distributions. Finally, we summarize our results in Section 5.


\section{Time of arrival formalism}
To measure the time of arrival of a free particle at a point $x$
one would: a) place a detector at $x$,  b) prepare the initial
state $|\psi\rangle$ of the particle at $t=0$, and then, c)  record
with a clock the time $t$ when the detector clicks. The value of
$t$ gives the {\sl toa} of the state $|\psi\rangle$ at $x$.
Repeating this procedure with identically prepared initial states,
one would get the probability distribution in times of arrival at
$x$. Of course, the results would depend on the initial state
chosen, which stores all the information regarding the initial
distribution in positions and momenta of the particle.

To determine the effect on these times of climbing (or tunneling
through) a potential barrier, one would simply put the barrier in
between the detector and the initial state, and then record the new
times of arrival. With an initial state identical to that prepared
for the free case, any difference in the probability distributions
should be an effect of the barrier. Several questions can be
investigated by changing the properties of the barrier: its height
or width if it is rectangular, even its very form.  This has been
explicitly done in the two photon experiments at Berkeley, by
putting alternatively a mirror and an ordinary glass in the path of
one of the photons. It would also be of interest to analyze the
dependence on $x$ when $V(x)>0$, i.e. with the detector within the
range of the interaction.

In classical mechanics particles move along the trajectories
$H(q,p)=$const. as $t$ increases. This allows to work out   $t_x$,
the time of arrival at the point $q(t)=x$, by identifying the point
$(q,p)$ of phase space where the particle is at (say) $t=0$,  and
then following the trajectory that passes by it, up to the arrival
at $x$. The mathematical translation of this procedure is given by
the equation of time:
\begin{equation}
t_x(q,p)=\mbox{sign}(p) \sqrt{\frac{m}{2}}\int_{q}^{x}\frac{dq'}
      {\sqrt{H(q,p)-V(q')}}\label{j5}
\end{equation}
that is discussed at length in many textbooks.

For free particles Eq. (\ref{j5}) gives $t_x^0(q,p)=m (x-q)/p$
that, in spite of its seemingly simple and harmless appearance,
presents some problems in its
quantization~\cite{Aharonov1,Tate,Juan}. First of all, it requires
symmetrization:
\begin{equation}
\hat{t}_x^0(\hat{q},\hat{p})=m (\frac{x}{\hat{p}}- \frac{1}{2}
\{\hat{q},\frac{1}{\hat{p}}\}_+)=-e^{-i\hat{p}x}\sqrt{\frac{m}{\hat{p}}}\hat{q}\sqrt{\frac{m}{\hat{p}}}e^{i\hat{p}x},
\label{jj}
\end{equation}
As is well known, the eigenstates of this operator $|t x s
0\rangle$ in the momentum and energy representations can be given
as ($\hbar =1$)
\begin{equation}
\langle p|t x s 0\rangle=\theta(sp) \sqrt{\frac{p}{2 \pi m}}\exp(i
\frac{p^2}{2m}t-i p x)\label{k155}
\end{equation}
where we use $s=r$ for right-movers ($p>0$), and $s=l$ for left
movers ($p<0$.) The indexlabel $0$ stands for free case. Finally,
the argument $sp$ of the step function that appears in the momentum
representation is $+p$ for $s=r$, and $-p$ for $s=l$. The
degeneracy of the energy with respect to the sign of the moment is
explicitly shown by means of the label $s$ in the energy
representation
\begin{equation}
 \langle E s' 0|t x s
0\rangle=\delta_{s's}\frac{1}{\sqrt{2\pi}} \exp(i E t-i s
\sqrt{2mE}x) \label{k15}
\end{equation}
where the $s$ in the exponent of the rhs stands for $+1$ for $s=r$
and $-1$ for $s=l$. These eigenstates are not orthogonal, which in
the past gave rise to serious doubts about their physical meaning.
The origin of this problem can be traced back to the fact that
(\ref{jj}) is not self-adjoint, that is $\langle\varphi|\hat{t}_x^0
\psi\rangle \neq \langle\hat{t}_x^0 \varphi| \psi\rangle$.  This
was proved by Pauli~\cite{Pauli} long time ago and is due to the
lower bound on the energy spectrum. The problem emerges as soon as
one attempts integration by parts in the energy representation.
Ref.~\cite{Muga6} contains a recent illuminating review of these
and other related questions.

A close inspection of (\ref{jj}) reveals that it still needs of
some prescription to deal with $p=0$, in order to produce a
self-adjoint operator with orthogonal eigenfunctions. When
$p\rightarrow 0^+$ the $\hat{t}_x^0$ eigenstates span the subspace
$\{|E\, r \rangle\}$, that comes abruptly to an end  at $p=0$, and
lacks the continuation to negative values of $p$ necessary to get
orthogonality (in other words, a delta function coming from the
integration over $p$ would require $p\in (-\infty,+\infty)$, but
actually $p\in (0^+,+\infty)$ alone). The way out~\cite{Tate} is to
stretch analytically the interval $p=(0,\epsilon)$ to
$p=(-\infty,\epsilon)$ in such a way that the new values of $p$
cover  all the real line. The same problem, with the same solution,
occurs when $p\rightarrow 0^-$. Observe that the attempt to ``glue"
$\{|E\, r\rangle\}$ to $\{|E\, l\rangle\}$ in a neigbourhood of
$p=0$ is misleading, because it would connect with continuity
states belonging to macroscopically different experimental setups
(those moving from the far left to the far right with small but
positive momentum, with those moving from the far right to the far
left with small but negative momentum). In fact, there is no
discontinuity in the eigenstates $|p\rangle$ at $p=0$ while, at the
same time, $\{|E\, r\rangle\}$ can not be connected continuously to
$\{|E\, l\rangle\}$. This is a consequence of the different
physical meaning of these two classes of states, given
mathematically by different kinds of boundary conditions. The
correspondence principle maps classical trajectories to energy
eigenstates, but does not restrict the momentum eigenstates in the
same way.

The measurement problem associated to the {\sl toa} operator can be
solved more simply by the use of a Positive Operator Valued Measure
(POV) , that only requires the hermiticity of $\hat{t}_x^0$ (i.e.
$\hat{t}_x^0={(\hat{t}_x^0)^*}^\top$). Here, instead of a Projector
Valued spectral decomposition of the identity operator, one has the
POV
\begin{equation}
P_x(t_1,t_2)= \sum_s \int_1^2 dt |t x s 0\rangle \langle t x s 0|
\label{j6}
\end{equation}
where $P(1,2)^2\neq P(1,2)$ because $|t x s 0\rangle \langle t x s
0|$ is not a projector, as the states are not orthogonal, but where
the limit as $t\rightarrow \infty$ of $P(-t,+t)$ is the identity.
The attained time operator is no longer sharp, but is well suited
for measurement. This solution has been implemented
in~\cite{Giannitrapani1}, and extensively analyzed in refs.
~\cite{Giannitrapani2,Toller1} and in the review~\cite{Muga6}.

In general, Eq.(\ref{j5}) can not be explicitly solved in the
presence of interaction. This is not a problem classically, but
makes hopeless any attempt to straight quantization.  Even for the
simple rectangular barrier that looks similar to a piece-wise
assembly of free cases, there is no successful way to employ
(\ref{j5}) directly. Here, the reason is not the algebraic
complexity of that expression, but the fact that it only applies
for values of  $x$ that are classically within the reach of
$(q,p)$, while in quantum mechanics all values of $x$ are possible.
Thus, neither quantum tunneling nor quantum reflection have a
classical analog in (\ref{j5}), (that gives complex numbers for
these cases). Lacking it, this equation is a troublesome starting
point for quantization. In spite of these problems, it
 arises as a critical time for the behaviour of particles in time
 dependent barriers, in the form of the so-called
 traversal time $\tau_T$: the time a particle takes to traverse the
barrier~\cite{Buttiker1} (which is the absolute value of
(\ref{j5})). As was indicated in the introduction, this time has
been compared, giving interesting information~\cite{Buttiker2}
about tunneling and related phenomena, with the ``dwell
time"~\cite{Smith} $\tau_d$, defined as the ratio of the number of
particles inside the barrier to the incident flux, and the ``phase
time"~\cite{Wigner} $\tau_\phi$, the time associated to the peak of
the propagating wave packet by the stationary phase approximation.

In this work we desist from the attempt of straight quantization of
the classical expression. Instead, we will construct the solution
to the interacting case in terms of the well known results that
apply to the free case. The aim is similar to what has been done in
formal scattering theory, where the Lippmann-Schwinger equations
give the complete Green function and the scattering amplitude in
terms of the free Green function and the potential. This procedure
also serves to get the energy eigenstates (stationary scattering
states) in terms of plane (or spherical) incoming or outgoing
waves.  Of particular interest for our problem is the existence of
the M\"oller operators $\Omega_\pm$ that connect the free particle
Hilbert spaces ${\cal H}_{in}$ and${\cal H}_{out}$, to the Hilbert
space $\cal H$ of the scattering and bound states. These operators
are only isometric in the presence of bound states, because the
correspondence between states in $\cal H$ and free states can not
be one to one. In this paper  we will consider only well behaved
potentials ($V(q)\geq 0\, \forall \, q$, $V(q)$ vanishes at the
spatial infinity, and has no valleys to avoid resonances), for
which the operators are unitary because there is one free state for
each scattering state. In this case, the intertwining relations $H
\Omega_\pm=\Omega_\pm H_0$ can be put in the form $H =\Omega_\pm
H_0\Omega^\dagger_\pm$. This is an useless relation as it stands,
but it prompts the following question: Is there a similar relation
for obtaining the time of arrival{\sl toa} in the presence of
interactions in terms of $\hat{t}_x^0$? Below, we construct a
positive answer, giving explicitly the generic solution for the
positive potentials we are considering. In the next section we will
apply our results to different cases of interest. There, we will
make the appropriate distinctions between incoming $(+)$ and
outgoing $(-)$ cases, that are necessary to predict the probability
distributions, but that do not affect the {\sl toa}, or its
eigenstates, when there is a unique free particle Hilbert space,
i.e. when ${\cal H}_{in}={\cal H}_{out}$ as in our case.

We now come back to phase space and recall that the classical
canonical transformations $\bar q=\bar q (q,p)$, $\bar p=\bar
p(q,p)$ can be defined implicitly by the use of auxiliary functions
$F,G,\bar F,\bar G$ in the following way:
\begin{eqnarray}
\bar F(\bar q,\bar p)& =&F(q,p)\\ \bar G(\bar q,\bar p)& =&G(q,p)
\label{k0}.
\end{eqnarray}
It is easy to work out the following relation among Poisson
brackets:
\begin{equation}
\{\bar F,\bar G\}_{\bar q \ \bar p}  \{\bar q,\bar p\}_{q \
p}=\{F,G\}_{q\ p} \label{k1}
\end{equation}
In these conditions, the transformation is canonical (i.e.  $\{\bar
q,\bar p\}_{q \ p}=1$) if and only if
\begin{equation}
\{\bar F,\bar G\}_{\bar q \ \bar p}=\{F,G\}_{q\ p} \label{k2}.
\end{equation}
 This relation has the additional property of fixing one of the four
 functions $F,G,\bar F,\bar G$, once the other three are given.
 We can choose $ F$ and $ G$ as the free particle Hamiltonian and time
 of arrival respectively. Then, if $\bar F$ is the complete Hamiltonian
 $H$, $\bar G$ will be the corresponding {\sl toa} $t_x$, and
 the above equation gives:$\{t_x,H\}=1$,
 whose solution along the classical trajectories is (\ref{j5}).

Canonical transformations were introduced by Dirac in quantum
mechanics by the use of unitary transformations $U$ ($U U^\dagger
=U^\dagger U=1$ ). If the operators $\bar {\hat q}, \bar {\hat p}$
are canonically transformed from $ {\hat q},  {\hat p}$ , then
there is a unitary transformation $U$ such that
\begin{eqnarray}
\bar {\hat q}&=&U \hat q U^\dagger \nonumber\\ \bar {\hat p}&=&U
\hat p  U^\dagger. \label{k4}
\end{eqnarray}
Then we can define implicitly quantum canonical transformations,
like the classical ones, by~\cite{Moshinsky3}
\begin{eqnarray}
\bar F(\bar {\hat q},\bar {\hat p})& =&U\bar F({\hat q},{\hat
p})U^\dagger =F({\hat q},{\hat p}) \nonumber \\ \bar G(\bar {\hat
q},\bar {\hat p})& =&U\bar G({\hat q},{\hat p})U^\dagger =G({\hat
q},{\hat p}) \label{k5}
\end{eqnarray}
where the last equality in each row is the definition of the barred
operators in terms of the unbarred ones, while the first equality
comes from the straight application of (\ref{k4}) to the l.h.s.
Being $U$ a unitary transformation, the spectra of the canonically
transformed operators have to coincide, that is:
\begin{eqnarray}
 \sigma (\bar {\hat q}) = \sigma ({\hat q}) = {\cal R},&
 \sigma (\bar {\hat p})= \sigma ({\hat p}) ={\cal R}\nonumber \\
 \sigma (\bar {\hat F})=\sigma ({\hat F}) ,& \sigma (\bar {\hat G})=
 \sigma ({\hat G})\label{k6}
\end{eqnarray}
where the second row stands because $F$ and $\bar F$, $G$ and $\bar
G$ are also unitarily related operators. We will now show how to
use the above relations to build the operator $\bar {\hat G}$ once
${\hat F},{\hat G}$ and $\bar {\hat F}$ are given. The only
restriction to our solution is that the operators  ${\hat F}$ and
$\bar {\hat F}$ are self-adjoint, so that  their eigenstates are
orthogonal and form a complete basis. The eigenstates of these
operators corresponding to the same eigenvalue $\lambda_f$ are:
\begin{equation}
\bar {\hat F}|{\bar f}\rangle = \lambda_f|{\bar f}\rangle,\,\,
{\hat F}|f \rangle = \lambda_f|f\rangle \label{k7}
\end{equation}
They form orthogonal and complete bases, i.e. they satisfy
\begin{eqnarray}
\langle f s|f's'\rangle=&\delta_{ss'}
\delta(\lambda_f-\lambda_f'),\, \,\sum_s \int_{\sigma (\lambda)}
d\lambda_f |f s\rangle \langle f s| &={\rm I\!I}\\ \langle \bar f
s|\bar f' s'\rangle=&\delta_{ss'} \delta(\lambda_f-\lambda_f'),\,\,
\sum_s \int_{\sigma (\lambda)} d\lambda_f |\bar f s\rangle
\langle\bar f s| &={\rm I\!I}\label{k8}
\end{eqnarray}
where we allow for some degeneracy (that has to be the same for
both ${\hat F}$ and $\bar {\hat F}$) labeled by $s$. We have also
assumed that $\lambda$ is continuous, while $s$ is a discrete
index. These assumptions could be changed straightforwardly if it
were necessary. Now, an operator $U$ satisfying the first row of
Eq.(\ref{k5}) can be given simply as:
\begin{equation}
U=\sum_s \int_{\sigma (\lambda)}  d\lambda_f |f s\rangle \langle
\bar f s| \label{k8}
\end{equation}
It is straightforward to verify that it is unitary. We can now
proceed to the sought for result: the definition of $\bar {\hat G}$
in terms of ${\hat G}$ using $U$, that is $\bar {\hat G}=U^\dagger
{\hat G} U$. The full fledged expression is
\begin{equation}
\bar {\hat G}({\hat q},{\hat p})=\sum_{ss'}
\int_{\sigma({\lambda})} d\lambda_f d\lambda_{f'}|\bar f s\rangle
\, \langle f s|{\hat G}({\hat q},{\hat p})|f's'\rangle \, \langle
\bar f 's'| \label{k9}
\end{equation}
that constitutes our main result in the quantum canonical
formalism. We will now apply it to the case where $F$ is the free
Hamiltonian $H_0$, $\bar F$ the complete Hamiltonian $H$: $\bar
F(\bar q,\bar p)=\bar p^2 /2 m +V(\bar q)$, and $G$ the time of
arrival of the free particle Eq.(\ref{jj}). Then we have
\begin{equation}
U_{(\pm)}=\sum_s \int_0^{\infty} dE |E s 0\rangle \langle E s
(\pm)| \label{k10}
\end{equation}
where the label $0$ in $|E s 0\rangle$ indicates that it is an
eigenstate of $H_0$, its absence corresponding to the eigenstates
$|E s \rangle$ of the complete Hamiltonian. For notation
simplicity, we defer to  Section 4 (Eq. ({\ref{fin1})) any
reference to the incoming or outgoing $(\pm)$ character of the time
operator and states we are constructing, as it is unnecessary here.
After substitution in Eq.(\ref{k9}) we get for the {\sl toa} at $x$
in the presence of the potential $V$
\begin{equation}
 {\hat t}_x ({\hat q},{\hat p})=\sum_{ss'} \int_0^\infty  dE dE'
 |E s\rangle \,
\langle E s 0|{\hat t}_x^0({\hat q},{\hat p})|E' s' 0\rangle \,
\langle E's'| \label{k11}
\end{equation}
where ${\hat t}_x^0$ is the {\sl toa} for the free problem, as
given by (\ref{jj}). The bracket in the above expression involves
free quantities only and can be computed to give
\begin{eqnarray}
\langle E s 0|{\hat t}_{x=0}^0({\hat q},{\hat p})|E' s' 0\rangle
&=- \langle E s 0|\sqrt{\frac{m}{\hat p}}{\hat
q}\sqrt{\frac{m}{\hat p}}|E' s' 0\rangle \nonumber \\ &=-i
\delta_{ss'} \left(\frac{d}{dE} \delta(E-E')\right) \label{k12}
\end{eqnarray}
where we have used $\langle p|E s0\rangle=\sqrt{\frac{m}{p}}
\delta(p-s\sqrt{2 m E})$, and put $x=0$ for simplicity (we will
restore the $x$ dependence later on). Using the above expression in
(\ref{k11}), we get the operator form
\begin{equation}
{\hat t}_{x=0} ({\hat q},{\hat p})=-\frac{i}{2} \sum_s
\int_0^\infty dE \, \, |E s\rangle\left(\stackrel{\leftrightarrow}{
\frac{d}{dE}}\right) \langle E s| \label{k13}
\end{equation}
that allows the explicit evaluation of the {\sl toa} for all the
cases in which the solutions $\psi_{Es}(q)=\langle q|E s \rangle$
to the Schr\"odinger equation are known.

By solving the eigenvalue problem posed by (\ref{k13}), we would
get the spectrum and eigenfunctions of the {\sl toa}. Instead of
using that procedure, we will do it in all generality (within our
assumptions about the potential and Hilbert spaces) by using the
canonical transformation $U$ to get the answer from the spectrum
and eigenfunctions of the free case. From the positive operator
valued decomposition of Eq.(\ref{j6}) we can write
\begin{equation}
\hat{t}_{x=0}^0=\sum_s \int_{-\infty}^{+\infty} dt\, t \,|t s
0\rangle \langle t s 0| \label{k14}
\end{equation}
where we still keep $x=0$, and  the spectrum of $\hat{t}_x^0$ is
$R$. The states $|t s 0\rangle$ are the non orthogonal eigenstates
of this operator given in (\ref{k15}). When used in (\ref{k12}),
Eq. (\ref{k14}) leads to
\begin{eqnarray}
\langle E s0|\hat{t}_{x=0}^0|E' s'
0\rangle&=&\sum_{s''}\int_{-\infty}^{+\infty} dt\, t \, \langle E s
0|t s'' 0\rangle \langle t s'' 0|E' s' 0\rangle\nonumber \\
&=&\frac{1}{2\pi}\delta_{ss'} \int_{-\infty}^{+\infty} dt\, t
\,\exp(i(E-E')t) \label{k16}
\end{eqnarray}
Putting this expression in Eq.(\ref{k11}) we get the spectral
decomposition for the {\sl toa} at $x=0$ in the presence of the
potential $V(q)$:
\begin{equation}
\hat{t}_{x=0}(\hat{q},\hat{p})=\sum_s \int_{-\infty}^{+\infty} dt\,
t \, |t s\rangle \langle ts| \label{k17},
\end{equation}
where the {\sl toa} eigenstates are defined by
\begin{equation}
|t s\rangle=\frac{1}{\sqrt{2\pi}}\int_{0}^{+\infty} dE\,
\exp(iEt)|Es\rangle. \label{k18}
\end{equation}
These eigenstates are not orthogonal. By direct computation using
(\ref{k18}) one gets
\begin{eqnarray}
\langle ts|t's'\rangle=\frac{1}{2\pi}\delta_{ss'}\int_{0}^{+\infty}
dE\, \exp(-iE(t-t'-i\epsilon))&=&\frac{1}{2\pi}\frac{i\,
\delta_{ss'}}{t'-t+i\epsilon}\nonumber \\ \sum_s
\int_{-\infty}^{+\infty} dt\,  |t s\rangle \langle ts|=\sum_s
\int_{0}^{+\infty} dE\,
 |E s\rangle \langle E s|&=&{\rm I\!I} \label{k19}
\end{eqnarray}
A set of relations defining the same kind of positive operator
valued measure that appears in the free case~\cite{Giannitrapani1}.
We can now restore the dependence of the different quantities on
the detector position $x$. Restarting from Eq.(\ref{k12}), but this
time keeping the term $m x/\hat{p}$ in the brackets, we arrive to
\begin{equation}
\hat{t}_x(\hat{q},\hat{p})=\sum_s \int_{-\infty}^{+\infty} dt\, t
\, |t x s\rangle \langle t x s| \label{k20},
\end{equation}
with the position dependent eigenstates given by
\begin{equation}
|t x s\rangle=\frac{1}{\sqrt{2\pi}}\int_{0}^{+\infty} dE\,
\exp(iEt- i s  x\sqrt{2mE})|Es\rangle. \label{k21}
\end{equation}
where the $s$ in the exponent of the rhs stands for $+1$ when
$s=r$, or $-1$ when $s=l$. We are now provided with the tools
necessary for a physical interpretation. Given an arbitrary state
$\psi$ at $t=0$, its time of arrival at a position $x$ has to be,
according to (\ref{k20}),
 \begin{equation}
\langle\psi|\hat{t}_x|\psi\rangle=\sum_s \int_{-\infty}^{+\infty}
dt\, t \, |\langle t x s| \psi\rangle|^2,\label{k22}
\end{equation}
with the standard interpretation of
\begin{equation}
\sum_s|\langle t x s| \psi\rangle|^2\label{k23}
\end{equation}
as the probability density that the state $|\psi\rangle$ arrives at
the detector placed at $x$ in a time $t$.

\section{Classical trajectories and energy eigenstates}
In three space dimensions, the free particle Hamiltonian $H_0$ is
independent of the position and of the direction of motion as a
consequence of the homogeneity and isotropy of space. In the one
dimensional case we are considering in this paper, $H_0$ continues
to be independent of the position, while the only remainder of
isotropy is the degeneracy of the Hamiltonian in the sign of the
momentum. For a given energy, the particle can move to the right or
to the left, and occupy any position in space. With the passage of
time (from $-\infty$ to $+\infty$), the particle will successively
pass by all the positions in configuration space. A steady flow of
(say) one particle of energy $E$ per unit time coming from the left
- and necessarily going to the right at the same rate - will
correspond to a properly normalized right mover stationary state.
This applies classically in phase space, as well as quantum
mechanically in the Hilbert space, and is also valid for left
movers.

In the presence of interaction, given by a potential energy $V(q)$,
space is no longer homogeneous, and the removal of degeneracies is
more involved. This is the case for the potential barriers we are
interested in. To simplify the analysis, avoiding bound states and
resonances whose effect we study elsewhere,  we will only consider
here barriers with positive potential that vanishes sufficiently
fast at the spatial infinity. In addition, these barriers  will
have  at most a maximum of value $V$, and no minima. Two textbook
barriers of this kind are the square barrier $V(q)=V$ if $0<q< a$,
else $V(q)=0$, and the well behaved barrier $V(q)=V/\cosh^2(q/d)$,
whose energy spectrum and eigenstates are well known. They can be
used to exemplify the material discussed in this section.

The classical trajectories associated to these barriers are simple
to compute, they are sketched in the phase space plane $(q,p)$ in
Fig.1, where the arrows on the trajectories indicate the direction
of motion. A main feature clear in both, the upper and lower
drawings, is the symmetry $(q,p)\leftrightarrow(q,-p)$, which comes
from the degeneracy in sign$(p)$. The symmetry under
$(q,p)\leftrightarrow(-q, p)$ is due to the spatial symmetry of the
potential chosen, and is largely irrelevant for the present
analysis. Another remarkable feature is given by the two symmetric
curves crossing the origin. Consider first in the upper figure the
curve that we call $p_V$ of equation $p=p_V(q)$, where $p_V(q)=
$sign$(q) \sqrt{2 m (V-V(q))}$. It divides the plane into two
regions. All the trajectories incoming to the barrier from the
left, and only them, are above $p_V$. These trajectories share the
property that as $t\rightarrow -\infty, (q,p)\rightarrow
(-\infty,\sqrt{2 m E})$, and split into transmitted trajectories in
the case when $E> V$ and reflected trajectories  when $E< V$. Below
$p_V$, we have the physically equivalent set of the trajectories
incoming to the barrier from the right and only them, verifying
that as $t\rightarrow -\infty, (q,p)\rightarrow (\infty,-\sqrt{2 m
E})$. As above, there are also transmission or reflection in this
region. The important thing is that the curve $p=p_V(q)$ serves to
separate the phase space into two distinct (physically equivalent)
regions in terms of  the asymptotic initial condition at
$t\rightarrow -\infty$. That is, above $p_V$ there are incoming
right-movers and incoming left-movers below it.

We could also make a partition of the phase space as shown in the
lower part of Fig.1, using the curve $p=-p_V(q)$. Above it, we find
all the outgoing right-movers; below it, all the outgoing
left-movers. There is also transmission or reflection depending on
whether $E> V$ or not. The partition of phase space is here in
terms of the final asymptotic condition for $t\rightarrow+\infty$:
above $-p_V$ there are right-movers ($(q,p)\rightarrow
(\infty,\sqrt{2 m E})$), and left-movers ($(q,p)\rightarrow
(-\infty,-\sqrt{2 m E})$) below it. Finally, we can count particle
trajectories
 in two alternative ways, as incoming or as outgoing trajectories,
according to the value of the asymptotic momentum at $t \rightarrow
- \infty$ or at $t \rightarrow +\infty$ respectively. For each way
there will be right and left movers according to the asymptotic
sign of $p$. The usefulness of a specific counting procedure will
be dictated by the objective pursued. For instance, to investigate
properties of the outgoing particles collected to the right (left)
of the barrier, we have to select from the set of trajectories
above (below) $-p_V$, a selection in terms of $p_V$ being useless
to this end.

In quantum mechanics one finds energy eigenstates in the Hilbert
space instead of the classical trajectories in phase space. A
particle will be described by a wave packet $\psi$, a suitable
superposition of energy eigenstates, that evolves according to the
Schr\"{o}dinger equation. If at some instant $t=0$ the state of the
particle is $|\psi\rangle$, the state at an arbitrary time can be
written, using completeness of the energy eigenstates, as
\begin{equation}
|\psi(t)\rangle=\exp(-i H t) |\psi\rangle= \sum_s \int dE \exp(-i E
t)|E\, s\, (\pm)\rangle\langle E\, s\, (\pm) |\psi\rangle
\label{a1}
\end{equation}
where $s=r,l$ gives the (asymptotic) sign of the momentum,
necessary to remove the degeneracy of the energy eigenstates, that
appears because $E$ is independent of $s$. The signs $(+)$ or $(-)$
indicate that the decomposition of the identity has been made in
terms of the corresponding energy eigenstates: $|E\, s\,
(\pm)\rangle=\Omega_{\pm}\, |E\, s\, 0\rangle$. These are the
quantum analogs of the classical trajectories.

We now analyze in more detail the two mathematically equivalent
ways for counting the energy eigenstates. As we said, the ($+$)
states represent a steady flow of particles going into the barrier,
corresponding to the partition of classical trajectories according
to $p_V$. These states can be expanded in terms of the energy
eigenstates $|E s {(+)}\rangle$. The label $s=r,l$ would correspond
to the sign of the momentum in the free case. Here, the states with
$s=r$ ($s=l$) are initial right-movers (left-movers). Observe that
in the presence of the potential the condition $E> V$ no longer
guarantees transmission, nor $E<V$ reflection. The $|E\, s\, {(+)}
\rangle$ states are widely used tools for studying the effects of
quantum barriers. The usual procedure is to prepare an incoming
right mover  $\langle q|E\, r\, 0\rangle= e^{i p q}$, and determine
the coefficients of transmission and reflection with the help of
the matching conditions. For each particle in the scattering state
$|E\, r\, {(+)} \rangle$, these coefficients give its probability
amplitude of finally being transmitted to the other side of the
barrier, or alternatively, of being reflected, bouncing back to the
left. Consider the case of a barrier contained in the range
$(0,a)$. The {($+$)} states can be written in configuration space
representation as:
\begin{eqnarray}
\langle q | E r {(+)} \rangle& =& \theta(-q) (e^{i p q} + R(p)
e^{-i p q})+ \theta(q) \theta(a-q) \chi_{r}(p,q)+\theta(q-a) T(p)
e^{i p q} \label{l1}\\
 \langle q | E l {(+)} \rangle& =& \theta(-q)
T(p) e^{-i p q}+ \theta(q) \theta(a-q) \chi_{l}(p,q)+\theta(q-a)
(e^{-i p q} + R(p) e^{i p q})\label{l2}
\end{eqnarray}
where $p=\sqrt{2 m E}$, the $\chi_{r,l}$ solve the
 Schr\"{o}dinger equation in the barrier region, and $R$ and $T$ are
 given by the matching conditions at $q=0$ and $q=a$. A (not displayed)
 overall normalization factor $\sqrt{m/2\pi p}$ will insure a steady
incoming flux of one particle per unit time.

The alternative way of counting energy eigenstates is in terms of
the {($-$)} states. These are the energy eigenstates that emerge
from the interaction region towards the spatial infinity.  The $|E
s {(-)}  \rangle$ are states going out from the barrier towards the
right ($r$), or towards the left ($l$). They correspond to the
classical partition according to $- p_V$. They describe a post
selection: that of the states that will be recorded with certainty
by the appropriate detectors placed to the right ($r$ case), or to
the left ($l$ case), of the barrier. By normalizing like the
{($+$)} states, there will be one record, i.e. one outgoing
particle per unit time in the steady flow corresponding to each of
these states. We can give them with the same notation as before:
\begin{eqnarray}
\langle q | E r {(-)} \rangle& =& \theta(-q)T(p) e^{i pq}+
\theta(q) \theta(a-q) \chi_{r}(p,q)+\theta(q-a)(e^{i pq} + R(p)
e^{-i pq}) \label{l3}\\
 \langle q | E l {(-)} \rangle& =& \theta(-q) (e^{-i pq}
+ R(p) e^{i pq})+ \theta(q) \theta(a-q) \chi_{l}(p,q)+\theta(q-a)
T(p) e^{-i pq}\label{l4}
 \end{eqnarray}

We said above that the {($+$)} and {($-$)} states give
mathematically equivalent decompositions of the unity ${\rm I\!I}$
in terms of the energy eigenstates and their degeneracies. In fact:
\begin{equation}
\sum_{s=r,l} \int dE |E s {(+)}\rangle \langle E s {(+)} |=
\sum_{s=r,l} \int dE |E s {(-)} \rangle \langle E s {(-)}|={\rm
I\!I}\label{l5}
\end{equation}
that was used in (\ref{a1}). As already noted, the physical meaning
of these states is very different. It is well known the form in
which the {(+)} states are used in three space dimensions to
describe the stationary scattering states, and to get the
scattering cross sections from its asymptotic form at
$|\vec{q}|\rightarrow\infty$. There, $\langle \vec{q}|\vec{p}
{(+)}\rangle\sim e^{i \vec{p}\vec{q}} + f(p,\Omega)\frac{e^{i p
q}}{q}$ and $(d\sigma/d\Omega)=|f(p,\Omega)|^2$, with $\vec{p}$ the
asymptotic incoming momentum and $\Omega$ the angle
$\widehat{(\vec{p},\vec{q})}$. In one dimension, the direction
given by $\Omega$ reduces to $r$ or $l$, and the scattering
amplitude $f(p,\Omega)$ to the appropriate transmission $T$ and
reflection $R$ coefficients. Less used, but also well known is the
association of the ($-$) states to a post selection of the final
states. For instance, $|\vec{p} {(-)} \rangle$ is the state that
comes out from the interaction region becoming asymptotically (at
$t\rightarrow\infty$)  the free state $|\vec{p} \rangle$.  There
are several different cases possible. They can be described by
means of projectors, a formulation that we will use later on.
Define
\begin{equation}
\Pi(E s {(+)})=|E s {(+)}\rangle \langle E s {(+)} |\label{l6}
\end{equation}
\begin{equation}
\Pi(E s {(-)})=|E s {(-)}\rangle \langle E s {(-)} |\label{l7}
\end{equation}
the projector $\Pi(E s {(+)})$ selects the states approaching the
barrier from the left ($s=r$), or from the right ($s=l$), with
energy $E$. In the same form, $\Pi(E s {(-)})$ selects the states
coming out from the right of the barrier ($s=r$), or from the left
($s=l$), with energy $E$.

Let $V$ be the maximum value of the potential energy. Classically,
there will be  full transmission through the barrier for $E>V$, and
complete reflection for $E<V$. As already said, this is not so in
the quantum case in which partial reflection and transmission will
take place for values of $k$ at which they can not occur
classically. In general, the wave packets will occupy regions of
space that are classically forbidden, giving rise to pure quantum
phenomena like tunneling, etc. It is the use of the above
projectors that will give us the control of these situations in the
Hilbert space, so that we can properly address the time of arrival
at a space point in the presence of barriers, mirrors, etc.

\section{Gaussian wave packets on  square barriers}
 This section is devoted to the theoretical predictions provided by
 the previous formalism  for the results of the experiments with
 potential barriers. We will analyze the case of transmission
 through a barrier in the first place, and then will turn our
 attention to the case of reflection, as both are aspects of the
 same problem. To fix notations, we will assume in what follows
 a square  barrier of height $V$ and width $a$, in the interval $(0,a)$,
  so that (\ref{l1}, \ref{l2}, \ref{l3}), and (\ref{l4}) are the energy
 eigenstates in the coordinate representation. We also assume that the
 detector is placed to the right of the barrier (that is, $x>a$), and
 that the initial state at $t=0$ is given by a Gaussian wave packet of
 width $\Delta q=\delta$, centered at $q_0<0$, with mean momentum
 $p_0>0$. The wave packet in configuration space and its Fourier
 transform are:
\begin{eqnarray}
\langle q|\psi\rangle& =&(\frac{1}{2 \pi \delta^2})^{(1/4)}\,
e^{-\delta^2 p_0^2}\,\, e^{-(\frac{q-q_0}{2\delta}- i \delta
p_0)^2}\label{l8}\\
 \langle p|\psi\rangle& =&(\frac{2\delta^2}{\pi
})^{(1/4)}\, e^{-\delta^2 (p-p_0)^2-ipq_0}\;\;\;\;(=
\tilde{\psi}(p)) \nonumber
\end{eqnarray}
For appropriate values of $q_0,p_0$ and $\delta$,  such that
$p_0\delta >>1$ and $\delta<<|q_0|$, almost all the packet is
initially at the left of the origin and moving with positive
momentum towards the barrier. We use this simplifying assumption in
our qualitative arguments, and in the intuitive description of the
process given below (indicated in the formulas by the use of
$\simeq$ instead of $=$), while working with  the full expression
(\ref{l8}) wherever necessary in the numerical calculations.

Our experiment differs from the usual ones in that it involves a
post selection. Usually, one prepares a well defined initial state
and observes its distribution into the possible final states. This
gives the probabilities of these outcomes. Here, we select the
final state of the particles coming out from the barrier to the
detector. The procedure is to collect all the particles in this
state, registering their times of arrival. From these records we
construct the conditional probability~\cite{Steinberg2} in time we
are interested in (i.e. in the {\sl toa}'s of {\sl all} the
particles gathered at the detector {\sl and only of them}). Given
the state $|\psi\rangle$, that of a particle placed at the left of
the barrier at $t=0$ as in (\ref{l8}), it is only the component
${\cal P}_{(-)}^r|\psi\rangle= \int dE \,\,\Pi(E r
{(-)})\,|\psi\rangle$ that will come out to the right of the
barrier (c.f. Eq.(\ref{l7})), arriving at $x>a$ in some time $t$.
Our formalism gives for the probability distribution in these times
of arrival
\begin{equation}
P_{(-)}^r(t|x)=\frac{1}{{N^r_{(-)}}^2}\sum_s|\langle t x s|{\cal
P}_{(-)}^r|\psi\rangle|^2 \label{prob}
\end{equation}
with the normalization coefficient $N^r_{(-)}$ given in the
equation (\ref{l14}) below.

We can now work out explicitly the probability amplitude for a
particle initially in the state $\psi$ to arrive at $x>a$ in a time
$t$. It will be given by:
\begin{equation}
\langle t x s|{\cal P}_{(-)}^r|\psi\rangle= \int dE \,\,\langle t x
s|\, E r {(-)} \rangle \langle \, E r {(-)} |\psi\rangle\label{l9}
\end{equation}
There are two factors in the integrand of the rhs. The first one is
obtained from the Eq. (\ref{k21}) of section 2:
\begin{equation}
\langle t x s|E r {(-)} \rangle=\frac{\delta_{sr}}{\sqrt{2\pi}}\,\,
\exp(-i E t+i  x \sqrt{2 m E})\label{l10}
\end{equation}
The second factor can be computed using the solution (\ref{l4}) to
the Schr\"odinger equation:
\begin{equation}
 \langle E r {(-)} |\psi\rangle =\int dq\,\,\ \langle
E r {(-)} |q\rangle \,  \langle q|\psi\rangle \label{l11}
\end{equation}
We said before that $\langle q|\psi\rangle$ practically vanishes
for $q>0$. Thence, the main contribution to the integral comes from
$q<0$, where $\langle q| E r {(-)} \rangle =\sqrt{m/2\pi p}
\,\,T(p) e^{i pq}$, with $p=\sqrt{2mE}$. In these conditions we
obtain
\begin{equation}
 \langle E r {(-)} |\psi\rangle\simeq \sqrt{\frac{m}{p}}
\,T^*(p)\,\tilde{\psi}(p)\label{l12}
\end{equation}
where $\tilde{\psi}(p)$ is the Fourier transform of $\langle
q|\psi\rangle$. Finally, the sought for probability amplitude is
\begin{equation}
\langle t x s|{\cal
P}_{(-)}^r|\psi\rangle\simeq\frac{\delta_{sr}}{\sqrt{2\pi}}\int
dE\,\sqrt{\frac{m}{p}}\,\,T^*(p)\,\,\tilde{\psi}(p)\,\,\exp(-iEt+ipx)
\label{l13}
\end{equation}
 Observe that $\langle t x l|{\cal P}_{(-)}^r|\psi\rangle= 0$
 as expected on physical grounds. The amplitude is not yet
 normalized, and has to be divided by the corresponding
 normalization factor $N^r_{(-)}$. In fact
\begin{equation}
{N^r_{(-)}}^2=\sum_s\int dt\,\,|\langle t x s|{\cal
P}_{(-)}^r|\psi\rangle|^2\simeq\int_0^{\infty}dp\,\,|T(p)|^2
|\tilde{\psi}(p)|^2\label{l14}
\end{equation}
i.e. ${N^r_{(-)}}^2$ is the probability that the particle be
transmitted, as corresponds to the case of conditional
probabilities we are considering. If the Gaussian chosen as initial
state is narrow enough in momentum space, and $p_0$ not too close
to $p_V=\sqrt{2mV}$, one can approximate it further to the
transmission probability
\begin{equation}
{N^r_{(-)}}^2\simeq |T(p_0)|^2 \int_0^{\infty}
dp\,\,|\tilde{\psi}(p)|^2\simeq|T(p_0)|^2\label{l15}
\end{equation}
as if all the transmission were at the mean momentum $p_0$. For
particles initially ($t=0$) in the state $\psi$, the average time
of arrival at a point $x$ at the other side of the barrier is:
\begin{equation}
t_x(\psi)=\frac{1}{{N^r_{(-)}}^2}\sum_s \int_{-\infty}^{+\infty} dt
\, t \, | \langle t x s|{\cal P}_{(-)}^r|\psi\rangle|^2 \label{l21}
\end{equation}
Any quantity related to the probability distribution
$P_{(-)}^r(t|x)$ can be easily obtained using (\ref{l13}) and
(\ref{l14}) or (\ref{l15}). Observe that the restriction to
gaussian wave packets is unnecessary for the validity of
(\ref{l13}) and (\ref{l14}). It is enough that $\psi(q)$ fulfills
the conditions given after (\ref{l8}).

In Fig. 2 we present the most probable {\sl toa} for a selected
range of barriers of various heights and sizes. To focus on the
effect of the barriers, we keep the same initial wave packet and
detector position in all the cases. What we obtain can be
summarized as retardation for $E>V$, and advancement for $E<V$. We
observe a close agreement between our results and the phase time
$\tau_\phi$ in the cases shown. This is because the stationary
phase method (that yields $\tau_\phi$) is a good mathematical
approximation to the true probability distribution in these cases
in which $\arg(T(p))$ varies slowly. We have also investigated
numerically (but do not show here) the cases $E\simeq V$ where the
validity of the approximation is uncertain. In fact, the values of
$\tau_\phi$ obtained in these cases run away from the true maximum
of $P_{(-)}^r(t|x)$, which turns out to be very sensitive to the
parameters $(p_0, \delta, V, a$) involved. In the cases shown, the
maxima give excellent approximations to the mean {\sl toa}
$t_x(\psi)$, even if the coincidence is not exact due to a small
amount of skewness present in the distributions.

 In the upper part
of Fig. 3, we show the probability distributions in {\sl toa}
$P^r_{(-)}(t|x)$ for three cases: free propagation, a low $V$
barrier, and transmission by quantum tunneling $E<V$. They look
quite similar, apart from the retardation or advancement pointed
out in Fig. 2. In fact, almost all of the many cases we have
investigated have a similar Gaussian look. It is only in the cases
in which $p_0 \delta>>1$ does not hold that new features may arise.
But these are very unrealistic one-particle wave packets, and the
results obtained from them unsound. We have analyzed in detail the
mechanism of advancement in quantum tunneling. Usually it is argued
that the higher momentum components are transmitted preferably
through the barrier, and that they propagate faster than the lower
momentum parts of the packet, being this the cause of advancement.
The argument is untenable for two reasons: First, advancement is
obtained even in the cases of packets whose spread is within a
range of momenta for which the transmission probabilities
$|T(p)|^2$ are constant (i.e. no preferred transmission for higher
$p$'s). Second, would the higher momenta be transmitted preferably,
then the transmitted average momentum should be shifted upwards,
and then, the transmitted packet should move faster than the
incident one! Of course this is physically untenable and false too.

We now turn to the time of arrival of states reflected from the
barrier. First of all, we place the detector at a position $y<0$ to
the left of the barrier. We keep the same barrier and the same
initial state. Now, it is ${\cal P}_{(-)}^l ={\rm I\!I} -{\cal
P}_{(-)}^r=\int dE \,\,\Pi(E l {(-)})$  that projects on the
subspace of the reflected states in which the particle is detected
at $y$ in some time $t$.  The corresponding probability amplitude
can be given borrowing Eq.({\ref{l9}}) from the transmission case,
\begin{equation}
\langle t y s|{\cal P}_{(-)}^l|\psi\rangle= \int dE \,\,\langle t y
s|\,  E l {(-)} \rangle \langle \,  E l {(-)}
|\psi\rangle\label{l16}
\end{equation}
There are again two factors on the right hand side. The first one
is  given by the formalism of section 2 as in (\ref{l10}). Note
that, according to (\ref{k21}), there will be a change in sign in
the second term of the exponent, as now $r\rightarrow l$. The
second factor , written as before, is
\begin{equation}
\langle  E l {(-)} |\psi\rangle =\int dq\,\, \langle E l {(-)}
|q\rangle \,\langle q|\psi\rangle \label{l17}
\end{equation}
 In the relevant region $q<0$, one has  from (\ref{l4}) that
 $\langle q| E l {(-)} \rangle=\sqrt{m/2\pi p}$ $( e^{-i pq}+R(p)
 e^{i pq})$, where $p=\sqrt{2 mE}$. We  now obtain the sum of two
 contributions (coming from 1 and $R$), giving
\begin{equation}
\langle  E l {(-)} |\psi\rangle \simeq \sqrt{\frac{m}{p}}
\,(\tilde{\psi}(-p)+R^*(p)\,\tilde{\psi}(p))\label{l18}
\end{equation}
As there is no way that a detector can discriminate between these
two contributions, they add in the amplitude, and may produce
potentially detectable interferences. However, for a properly
chosen initial wave packet with positive mean momentum such that
$p_0 \delta>>1$ or, in other words, with negligible negative
momentum tail, the first term in the sum on the rhs will
practically vanish. Dropping this term, we get the probability
amplitude as given by:
\begin{equation}
\langle t y s|{\cal
P}_{(-)}^l|\psi\rangle\simeq\frac{\delta_{sl}}{\sqrt{2\pi}}\int
dE\,\sqrt{\frac{m}{p}}\,\,R^*(p)\,\,\tilde{\psi}(p)\,\,\exp(-iEt-ipy)
\label{l19}
\end{equation}
which is a look-alike of the transmission case (\ref{l13}). The
normalization factor $N^l_{(-)}$  is also similar:
\begin{equation}
{N^l_{(-)}}^2=\sum_s\int dt\,\,|\langle t,y,s|{\cal
P}_{(-)}^l|\psi\rangle|^2\simeq\int_0^{\infty}dp\,\,|R(p)|^2
|\tilde{\psi}(p)|^2\simeq|R(p_0)|^2\label{l20}
\end{equation}
where the last approximation is  valid only when  the initial
 state $\psi(q)$ is narrow enough and, as for transmission,
$p_0$ not too close to $p_V$. We can now give the average time of
arrival at a point $y$ to the left of the barrier of the outgoing
particles that were in the initial state $\psi$ at $t=0$:
\begin{equation}
t_y(\psi)=\int_{-\infty}^{+\infty} dt \,t\, P_{(-)}^l(t|y) =
\frac{1}{{N^l_{(-)}}^2}\sum_s\int_{-\infty}^{+\infty} dt \, t \, |
\langle t y s|{\cal P}_{(-)}^l|\psi\rangle|^2 \label{l22}
\end{equation}
with a probability distribution $P_{(-)}^l(t|y)$ that is also
normalized to one. In the lower part of Fig. 3 we show the
probability distributions for reflection by the same potential
barriers of the upper part. When the barrier acts as a classical
mirror (the long dashes curve) the distribution is almost the same
as that of a free particle that would travel the same distance.
However, dramatic differences appear for purely quantum reflection
(i.e. when $E>V$), whenever the condition $p'_0 a = n \pi,\; n=
1,2,...$ (where $p'_0 =\sqrt{p_0^2 -k_v^2}$ is the momentum inside
the barrier) holds. In these cases, a two bump structure with a dip
in the middle appears in the distribution. The reason of this
behaviour can be found in the value of the reflection coefficient
\begin{equation}
R(p)= \frac{k_v^2 \sin(p' a)}{\sqrt{2 p^2 p'^2 +k_v^4 \sin^2(p'
a)}} \exp(i \pi - i \arctan(\frac{k_v^4}{4 p^2 p'^2}\tan(p'
a)))\label{reflection}
\end{equation}
which vanishes whenever the above condition is satisfied. The
proportionality between the reflection probability and $\sin^2(p'
a)$, that originates the dip, could also be used to explain the
bumps as due to reflection of the components of momenta greater and
lesser than $p_0$. Were this the case, the bumps should be
reflected with different velocities, one above and the other below
$p_0/m$! Of course, this does not hold. The actual mechanism of the
bumps is the spread of the incident wave packet in configuration
space (not in momentum space). The centre of the packet (placed
initially at $q=q_0$) is reflected with probability zero, while the
bumps are composed by the parts ahead and behind it (both carrying
main momentum $p_0$). Here, as in the case of tunneling, there is a
matching between the real and imaginary exponents in the
probability amplitude. Thus, while the participating momenta are in
the small range $p_0\pm \delta^{-1}$, due to the Gaussian shape of
the wave packet, the distances and times involved are macroscopic.
Therefore, the phases of the different contributions within this
range of $p$ are rapidly varying and their sum cancels out. The
only remaining survivors come from the stationary phases $p\simeq m
(x-q_0- \arg'(T(p)))/t$ or $p\simeq m (-y-q_0 -\arg'(R(p)))/t$.
These contributions have still to be weighted by the different
remaining factors, thus giving rise to the observed probability
distributions, in which $p$ is a function of $t$, and depends
parametrically on $x, (y)$ and $q_0$ according to the above
relations. To illustrate the result of this mechanism, it is enough
to point out the approximate relation $ P_{(-)}^l(t|y) \propto
 \sin^2(\sqrt{p_0(t)^2-k_v^2}\; a)
 \exp(-2 \delta^2 (p_0(t)-p_0)^2)$ with
 $p_0(t)=m(-y-q_0-\arg' R(p_0))/t$, as the main responsible of the
two bump structure.

We now recall that the {\sl toa} of incoming states can be
described with the same formalism used above. It is now necessary
to project on the subspace of incoming states using $\Pi(E s
{(+)})$. The physical picture here is a detector placed before the
barrier to intercept these states. There are two opposite incoming
directions. Mathematically, they are selected by the two projectors
$\Pi(E r{(+)})$ (for detecting right movers at the left of the
barrier), or $\Pi(E l {(+)})$ (for left movers at the right). We
choose as an example the first of these cases. The probability
amplitude in times of arrival at $y$ of the incoming states is:
\begin{equation}
\langle t y s|{\cal P}_{(+)}^r|\psi\rangle= \int dE \,\,\langle t y
s|\,  E r {(+)} \rangle \langle \,  E r {(+)}
|\psi\rangle\label{l23}
\end{equation}
Now, instead of (\ref{l18}), we have
\begin{equation}
\langle  E r {(+)} |\psi\rangle \simeq \sqrt{\frac{m}{p}}
\,(\tilde{\psi}(p)+R^*(p)\,\tilde{\psi}(-p))\label{l24}
\end{equation}
  As occurred for reflection, there are two potentially
interfering terms, due to the lack of discrimination between the
direct incoming component and the receding part of the reflected
component. However, only the first term will give a sizeable
contribution when the negative momentum tails are negligible. It is
worth noting that the direct component gives the same time of
arrival distribution that in the absence of barrier, as it should
be on physical grounds. In the stationary phase approximation, the
peak of the distribution is on the classical {\sl toa}: $t=m
(y-q_0)/p_0$, as expected. This also gives the meaning of the
negative {\sl toa}'s~\cite{Tate}: they are obtained when the
detector is placed to the left of the incoming right mover, so that
the particle would have arrived at it before $t=0$. Finally, we
note that the probability amplitude of incoming states at a point
$x$ to the right of the barrier would be proportional to the
receding part of the transmitted state.

We now restore the $(\pm)$ labeling that should be explicitly
included in the {\sl toa} formulation of Section 2, and that we
ignored from Eq. (\ref{k10}) on. By doing it, instead of the
equations (\ref{k20},\ref{k21}) at the end of that section we would
arrive to
\begin{equation}
\hat{t}_x(\pm)=\sum_s \int_{-\infty}^{+\infty} dt\, t \, |t x s
(\pm) \rangle \langle t x s (\pm)| \label{fin1},
\end{equation}
with the position dependent eigenstates given by
\begin{equation}
|t x s (\pm)\rangle=\frac{1}{\sqrt{2\pi}}\int_{0}^{+\infty} dE\,
\exp(iEt- i s  x\sqrt{2mE})|Es (\pm)\rangle. \label{k21}
\end{equation}
We thus see that many of the physical considerations made in this
section in order to obtain the probability distributions in {\sl
toa} are automatically taken into account by the formalism. Instead
of Eq. (\ref{prob}) we would get directly
\begin{equation}
P_{(-)}^r(t|x)=\frac{1}{{N^r_{(-)}}^2}\; |\langle t x r
(-)|\psi\rangle|^2 \label{prob2}
\end{equation}
and similar abridged (but fully equivalent) expressions in
analogous cases. We recall from a formal point of view, that
 there are two possible constructions of $t{(\pm)}$
starting from the free {\sl toa}, as there are two constructions of
the stationary scattering states $|E s (\pm)\rangle$ starting from
the free states $|E s 0\rangle$. Even if each of these, $t(+)$ or
$t(-)$, is associated to a kind of asymptotic state, {\sl in} or
{\sl out} respectively, nowhere in the construction have we imposed
or used any asymptotic value for $x, y, t$ or for any other
quantity in the problem. It is the finite extension of the barrier
the only responsible for the simple asymptotic looking of our
expressions. We used this to discuss the physical questions with a
minimum of burden.

\section{Conclusions}
We have worked out a formalism for obtaining the time of arrival at
a space point of particles that move through interacting
environments. Our construction follows a circuitous path; we desist
from first computing the classical {\sl toa} of the problem, and
then quantize it, a procedure that leads to a dead end. Instead, we
start from the quantum {\sl toa} of the free moving particle, and
then transform it canonically to the interacting case. This is
achieved by the use of the appropriate M\"oller operators, that we
employ in a nonstandard, but orthodox, setting. Usually, they are
introduced to connect the values (at $t=0$) of those free and
scattering states that will coincide asymptotically (at
$t\rightarrow\pm\infty$, depending on the case). Here, we first
derive them from the connection between the free and interacting
Hamiltonians, and then use them to connect the free and interacting
{\sl toa}'s, which solves our problem. For simplicity, we have only
addressed explicitly the cases in which the transformations are
unitary. The well known results from formal scattering theory pave
the way to the general cases that require isometric
transformations.

We have also performed a quite exhaustive (and a bit heavy)
analysis of the relation between classical trajectories and
scattering states. We found necessary to delve into the parallelism
between the partitions of classical phase space by means of {\sl
in} or {\sl out} trajectories on the one side, and the spans of the
Hilbert space by means of the $(+)$ or $(-)$ states on the other.
The right appreciation of the different issues involved in that
correspondence was necessary for a successful use of the formalism
previously developed. At the end, we reached the conclusion that
this formalism did incorporate from the outset these subtleties. In
fact, for the asymptotically vanishing potentials considered here
we obtain two different operators $t(+)$ and $t(-)$. They measure
the {\sl toa} at a space point of the incoming and outgoing states
respectively.

In the course of our numerical analysis we have detected that the
phase time $\tau_\phi$ gives a good approximation to the most
probable time of arrival. It provides a first estimate of the time
spent in the transmission or reflection, after subtracting the time
of free flight. We have found advancement of the transmitted wave
packet in the case of pure quantum tunneling. This is a long known
phenomenon first predicted by Hartman~\cite{Hartman}, and
experimentally evinced by the two photon experiments at
Berkeley~\cite{Steinberg3,Steinberg4}. We have found an unexpected
phenomenon for purely quantum reflection: the two bump structure
that appears when $p'_0 a=n \pi$. We have shown it only in one case
in Fig. 3, which by no means is the most crisp case. Neat double
humps appear in all the expected cases, the lesser the potential
barrier the finer the feature. With the insight provided by our
formalism, we have also proposed an explanation to this structure.
We think this feature, even if less spectacular than the
superluminal tunneling of photons, deserves experimental
confirmation. A minor modification of the two photon experiment
could serve for this purpose. It would be enough to place a quantum
mirror in the path of one of the entangled photons, and check for
the presence (or absence) of a two dip structure in the number of
coincidence counts.

\section*{Acknowledgements}
J. L. wishes to thank S. Brouard, V. Delgado, R. Sala, and
specially J. Muga, for many useful discussions. He  acknowledges
the important collaboration of R. S. Tate in previous analysis of
the {\sl toa}. Finally, J. J.  and J. L.  thank the La Laguna Group
for the invitation to participate in the workshop "Time in Quantum
Mechanics" held in June 1998.

\begin{figure}
\epsfxsize=16cm \epsfbox{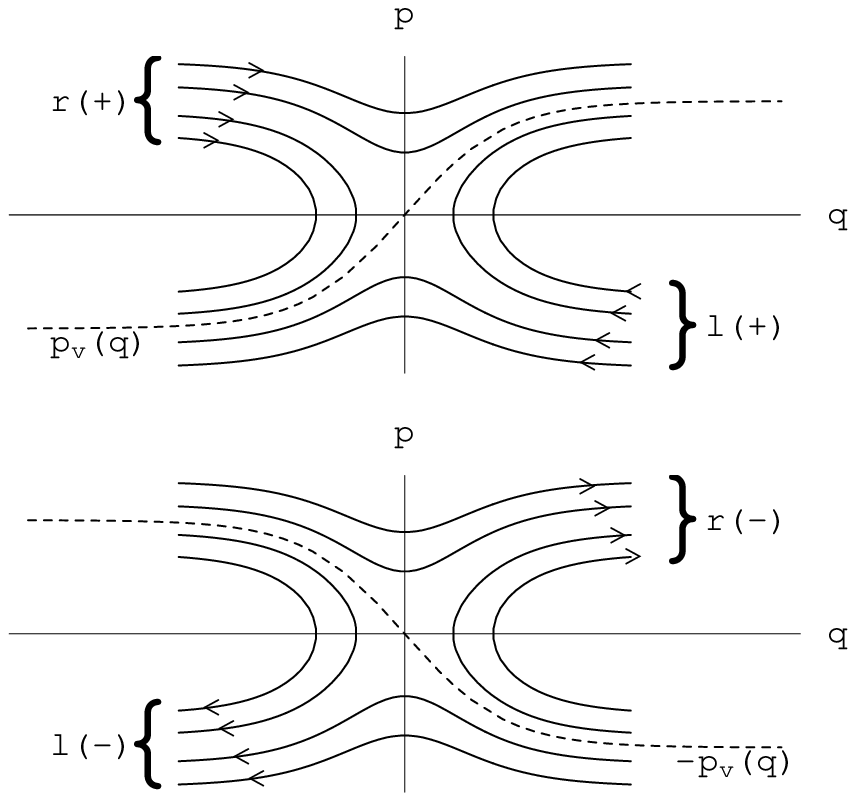}
 \caption{ Alternative partitions of the phase space plane $(q,p)$ in terms
  of the trajectories of  particles with Hamiltonian
 $H(q,p)=p^2/2 m +V/\cosh^2(q/d)$.
The trajectories of incoming right movers $r (+)$ are above the
 separatrix $p=p_V (q)$, and those of incoming left movers $l (+)$
 below it, both in the upper figure.
 The trajectories of outgoing right movers $r (-)$, and left movers
 $l (-)$ are respectively above and below $p=-p_V (q)$ in the lower figure.
   The notation is borrowed from that of the stationary
 scattering states}
\label{fig1}
\end{figure}
\begin{figure}
\epsfxsize=16cm \epsfbox{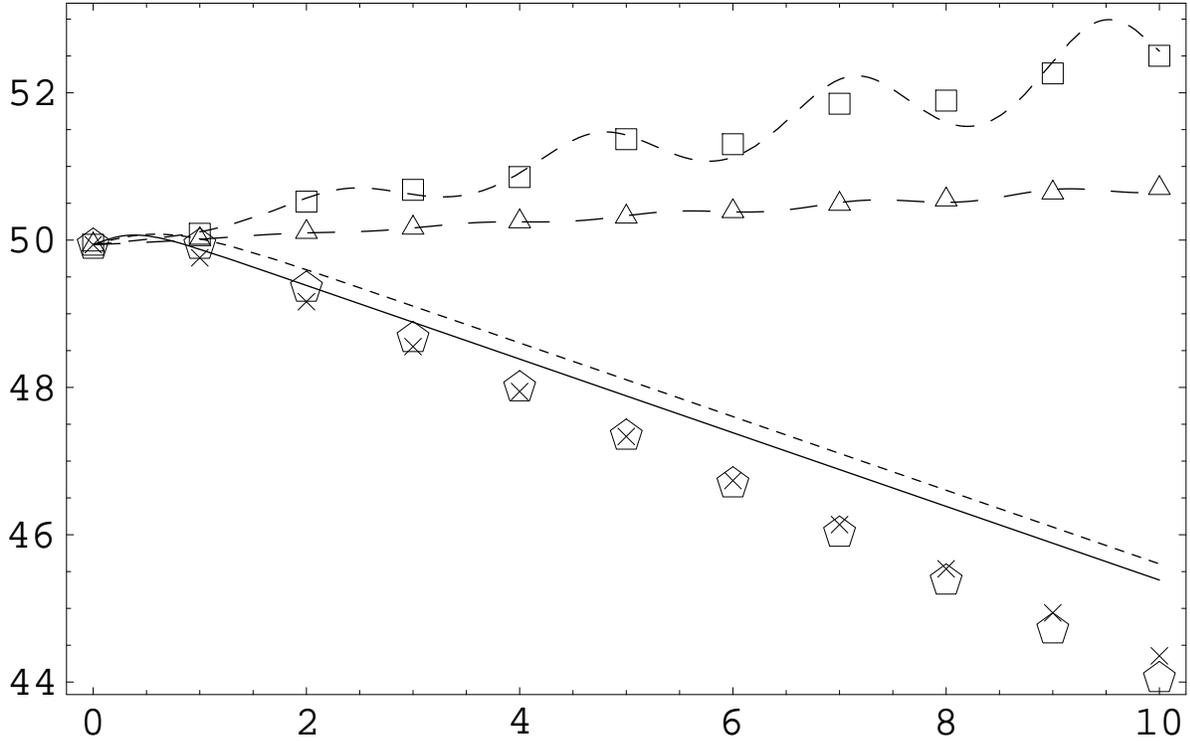}
 \caption{Most probable time of arrival and phase times for
 transmission through barriers of different heights and widths.
 Times are represented on the vertical axis against the barrier width
 ranging from 0 to 10. The initial wave packet (c.f. Eq. (\ref{l8})) has
 $q_0=-50,\, p_0=2,\, \delta=10$ and $m=1$. The detector is at $x=50$. These
 parameters are fixed, so that the classical {\sl toa} for free propagation
 is $50$ in all cases, (with our packet, we get the most probable {\sl toa}
 at $t_0=49.9377$
 quantum mechanically). We show our results for the most
  probable {\sl toa} with triangles, squares, pentagons, and crisscrosses,
corresponding to barrier heights of $V=0.5,\, 1.125,\, 3.125,$ and
 $4.5$ respectively. The
 curves represent ``$t_0+$ Wigner time delay" for these same heights,
 (being the solid one for $V=4.5$, etc). The plots neatly indicate
 retardation when transmission is classically possible,
 and advancement for pure quantum tunneling.}
\label{fig2}
\end{figure}
\begin{figure}
\epsfxsize=16cm \epsfbox{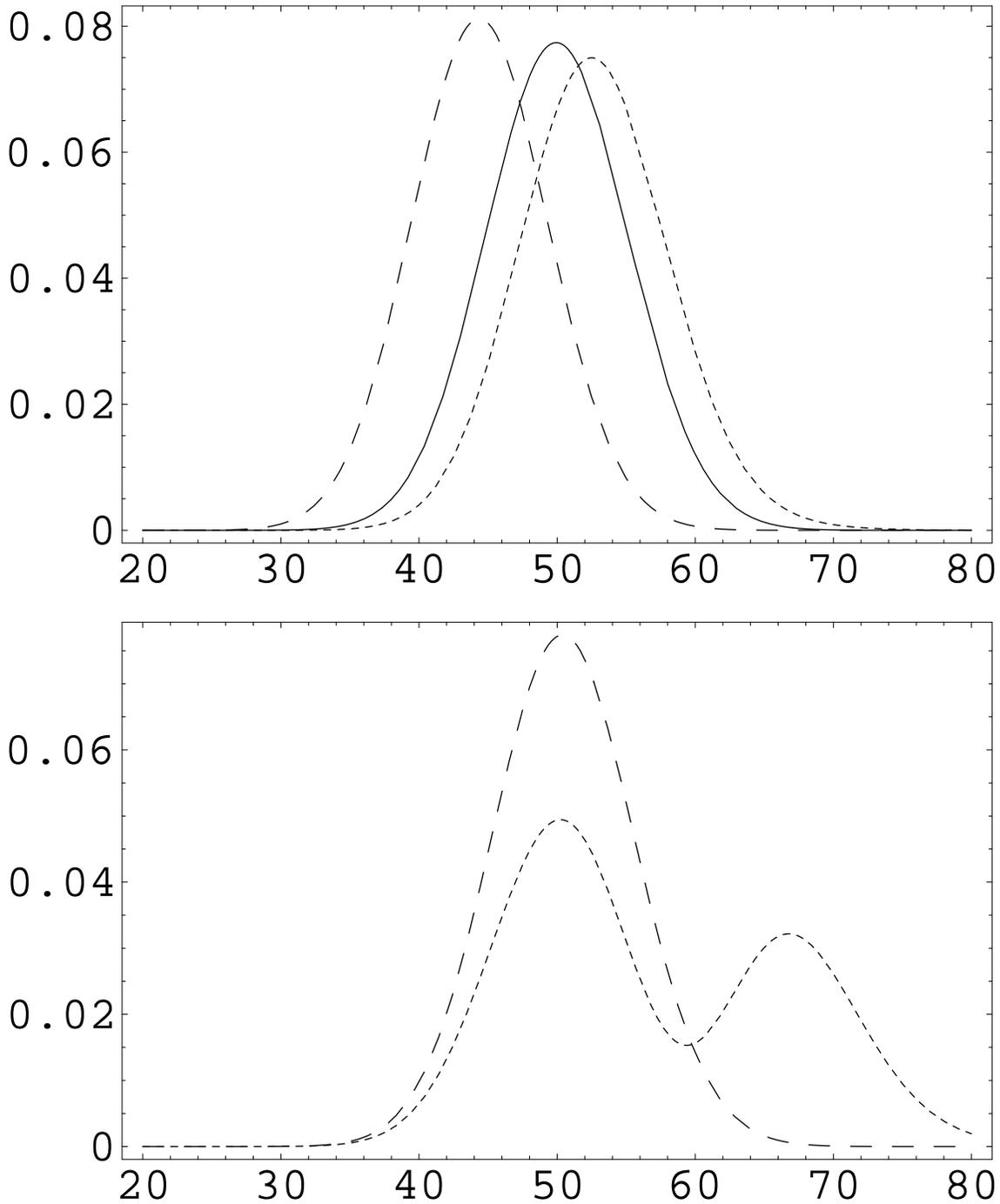}
 \caption{Probability distributions in times of arrival for
 transmission (above) and reflection (below), corresponding to
  equations (\ref{prob})
  and (\ref{l22}). The initial wave
 packet is  in all cases the same of Fig.(\ref{fig3}). The total
 length of the particle's path is always 100 units, so that the classical
free {\sl toa} is 50. The width of the barrier is $a=10$. The solid
line represents free transmission, the short dashes the cases with
$V=1.125$, and the long ones those with
 $V=4.5$. The two bump structure is a general result in the cases where
 there is no classical reflection as explained in the text.} \label{fig3}
\end{figure}

\end{document}